\documentclass[a4paper, 10 pt, twocolumn]{article}
\pdfoutput=1
\usepackage{cite}
\usepackage{algorithm}
\usepackage{textcomp}
\usepackage{graphicx}

\usepackage{amsmath,amsfonts,amssymb}
\usepackage{epsfig,dsfont,amssymb,amsthm,amsbsy,mathrsfs,color} 
\usepackage{dcolumn}   % needed for some tables
\usepackage{bm}        % for math
\usepackage{lineno,hyperref}  %for ref
\usepackage{cases}     %for equation 1(a)1(b)
\usepackage{multirow}  %for table
\usepackage{array}

\newcolumntype{Y}{>{\centering\arraybackslash}X}

\usepackage{threeparttable}
\usepackage{footnote}
\usepackage{xcolor}

\usepackage[]{algpseudocode} 
\usepackage{multirow}
\usepackage{dcolumn}
\usepackage{times}  
\usepackage{helvet}
\usepackage{courier}  
\PassOptionsToPackage{hyphens}{url}
\usepackage{graphicx}
\usepackage{graphicx} 
\usepackage[colorinlistoftodos]{todonotes}
\PassOptionsToPackage{colorlinks=true, allcolors=black}{hyperref}
\usepackage[acronym]{glossaries} %for abbv list
\usepackage{colortbl}
\usepackage{hhline}
\usepackage{changepage}
\usepackage{subcaption,siunitx,booktabs}
\usepackage{adjustbox}
\usepackage{pifont}%
\usepackage{emptypage}
\usepackage{tabularx}
\usepackage{threeparttable}

\begin{document}
\title{Basal Glucose Control in Type 1 Diabetes using Deep Reinforcement Learning: An \textit{In Silico} Validation}

\author{Taiyu~Zhu$^{1*}$, Kezhi~Li$^{1,2*}$, Pau~Herrero$^1$, Pantelis~Georgiou$^1$\\ 
$^1$ Department of Electronic and Electrical Engineering, Imperial College London\\
$^2$ Institute of Health Informatics, University College London\thanks{The work was supported by EPSRC EP/P00993X/1, names with $*$ have equal contribution. The paper is under review.}
}

\date{}
\maketitle

\begin{abstract}

People with Type 1 diabetes (T1D) require regular exogenous infusion of insulin to maintain their blood glucose concentration in a therapeutically adequate target range. Although the artificial pancreas and continuous glucose monitoring have been proven to be effective in achieving closed-loop control, significant challenges still remain due to the high complexity of glucose dynamics and limitations in the technology. In this work, we propose a novel deep reinforcement learning model for single-hormone (insulin) and dual-hormone (insulin and glucagon) delivery. In particular, the delivery strategies are developed by double Q-learning with dilated recurrent neural networks. For designing and testing purposes, the FDA-accepted UVA/Padova Type 1 simulator was employed. First, we performed long-term generalized training to obtain a population model. Then, this model was personalized with a small data-set of subject-specific data. \emph{In silico} results show that the single and dual-hormone delivery strategies achieve good glucose control when compared to a standard basal-bolus therapy with low-glucose insulin suspension. Specifically, in the adult cohort (n=10), percentage time in target range $[70, 180]$ mg/dL improved from $77.6\%$ to $80.9\%$ with single-hormone control, and to $85.6\%$ with dual-hormone control. In the adolescent cohort (n=10), percentage time in target range improved from $55.5\%$ to $65.9\% $ with single-hormone control, and to $78.8\%$ with dual-hormone control. In all scenarios, a significant decrease in hypoglycemia was observed. These results show that the use of deep reinforcement learning is a viable approach for closed-loop glucose control in T1D.
 
\end{abstract}
\noindent

\section{Introduction}
Diabetes is a chronic disease which affects millions of people worldwide. It is characterised by elevated blood glucose (BG) which in the long term can lead to complications such as cardiovascular disease, retinopathy and nephropathy. Its global prevalence rate has reached epidemic proportions, doubling in the last 20 years \cite{cho2018idf}. There are two main types of diabetes, Type 1 and Type 2. Type 2 diabetes is characterised by the body ineffectively using insulin and can be usually treated with lifestyle interventions and oral medication. Type 1 diabetes (T1D) however is distinguished by insufficient insulin production by the pancreatic $\beta$-cell and therefore requires exogenous insulin administration. The standard insulin replacement therapy for T1D includes a bolus of fast-acting insulin to compensate the fast glucose increase after meal ingestion, and a basal insulin delivery through an injection of slow-acting insulin to keep glucose levels within target range in fasting conditions. Alternatively, basal insulin can be delivered through continuous infusion using an insulin pump with fast-acting insulin. Although software tools such as bolus calculators exist to support people with T1D to self-administer insulin, they still fall short to achieve optimal glycemic control \cite{schmidt2012use}. Therefore, realising an automated system to deliver optimal insulin doses is one of the long-standing challenges in glucose management over the past decades \cite{quiroz2019evolution}.

Recent improvements in accuracy and reliability of continuous glucose monitoring (CGM) systems has allowed the development of a closed-loop insulin delivery system, also known as the artificial pancreas (AP), to automatically control BG levels in T1D \cite{kovatchev2019century}. An AP consists of, at least, a CGM sensor, a control algorithm, and an insulin pump. Additionally, some AP systems might also incorporate a glucagon pump to counter-regulate the action of insulin \cite{haidar2019insulin} and an activity monitor to quantify physical exercise \cite{deboer2017heart}. Glucose measurements are captured by the CGM device every five minutes and are sent to the control algorithm which calculates the corresponding dose of insulin aiming at maintaining glucose level in a target range, which is then delivered by the infusion device. To date, most existing AP approaches that have been evaluated in clinic have used a control engineering approach \cite{doyle2014closed}, and one uses artificial intelligence (fuzzy logic) \cite{atlas2010md}. In particular, two of them, the Medtronic 670G and the Tandem Control-IQ have reached the commercialisation stage. However, although these systems have been proven to improve glycemic control \cite{messer2018optimizing, forlenza2019successful}, challenges remain, and further work is needed to achieve optimal therapeutic targets. 
 
In recent years, powered by the large scale of available medical data and the rapid advances in computational power, machine learning, in particular deep learning, has increasingly been used in many healthcare applications that were out of reach in the past \cite{jiang-artificial2017}, especially in diagnostics and medical imaging \cite{Esteva-dermatologist2017,gargeya2017automated}.  

In the field of diabetes, the use of machine leaning has also attracted significant attention \cite{contreras2018artificial}. In particular, neural networks (NN) have achieved success in glucose forecasting \cite{perez-ArtiNN2010} (fully-connected neural networks), \cite{zhu2018deep,li2019glunet} (convolutional neural networks), \cite{Chen-DilatedRec2018,Li-CRNN2019,zhudilated2020} (recurrent neural networks (RNN)), and \cite{Bertachi-PreofBlo2018} (physiological-based networks). Of note, dilated RNN (DRNN) has performed particularly well in processing long-term dependencies and future glucose prediction \cite{chang-dilated2017,Chen-DilatedRec2018,zhudilated2020}. 
Recently, another technique under the spotlight in the field of automatic insulin delivery is reinforcement learning (RL) \cite{tejedor2020reinforcement}. RL is a machine learning framework for learning sequential decision-making tasks. Many healthcare problems, such as drug delivery, can be seen as closed-loop sequential action-selection problems, which is what RL focuses on \cite{mnih-human-level2015}. Unfortunately, the use of deep RL in healthcare has been limited by several practical issues. Unlike successful deep RL applications in the virtual world, such as Atari video-games or board-game Go \cite{mnih-human-level2015,silver2017mastering} where an agent dynamically interacts with a virtual environment,
performing such exploration on human subjects can be dangerous without proper safety constraints. Alternatively, deep RL algorithms can learn from existing collected data using experience replay. This process is called off-policy learning and plays an important role in practical RL algorithms. However, collecting the training data required is expensive and time consuming \cite{Artman-power2018}. Fortunately, an FDA-accepted T1D simulator developed in collaboration between the University of Virginia (US) and the University of Padova (Italy) is available for developing and evaluating insulin and glucagon delivery strategies \cite{Man-UVA/PADOVA2014}. 

In this paper, we explore, \textit{in silico}, the use of deep RL for closed-loop control of BG levels in T1D.
The paper is organized as follows. Section~\ref{sec:Method} describes the architecture and algorithms of the proposed deep RL framework for glucose control. The performance of the proposed method is evaluated in Section~\ref{sec:Experiments}. Section~\ref{sec:Discussion} compares the results with existing work and discusses the future work. Finally, we summarize the work in Section~\ref{sec:Conclusion}.

\section{Methodology}
\label{sec:Method}
In this section, we state the problem of basal blood glucose closed-control in terms of deep RL. Then, we introduce a two-step framework, adapted from transfer learning, to develop, \textit{in silico}, single and dual-hormone glucose controllers to be potentially used in clinical practice. 

 In particular, a deep Q-learning model \cite{mnih-human-level2015} is employed to optimize insulin and glucagon delivery. Insulin and glucagon dose deliveries are treated as actions ($a$) taken by a stochastic policy, glycemic outcomes (e.g. percentage time in glucose target) are considered as rewards ($r$), and physiological variables are seen as states ($s$). A deep neural network (DNN) is used as a non-linear function approximator to estimate action-values, also referred to as a deep Q-network (DQN). Unlike previous methods to control AP systems using conventional RL, no prior knowledge of the glucose-insulin-glucagon metabolism is needed. Instead, a stack of dilated recurrent layers is used for processing multi-dimensional time series data. Because of its enlarged receptive field, it is able to capture the complexity of glucose-insulin-glucagon dynamics, as shown in our previous study \cite{Chen-DilatedRec2018,zhudilated2020}. Section \ref{sec:DQN_selection} in the Appendix explains how the DRNN model was selected over other neural network architectures.

\begin{figure*}[t]
\centering
 \includegraphics[width=0.8\textwidth]{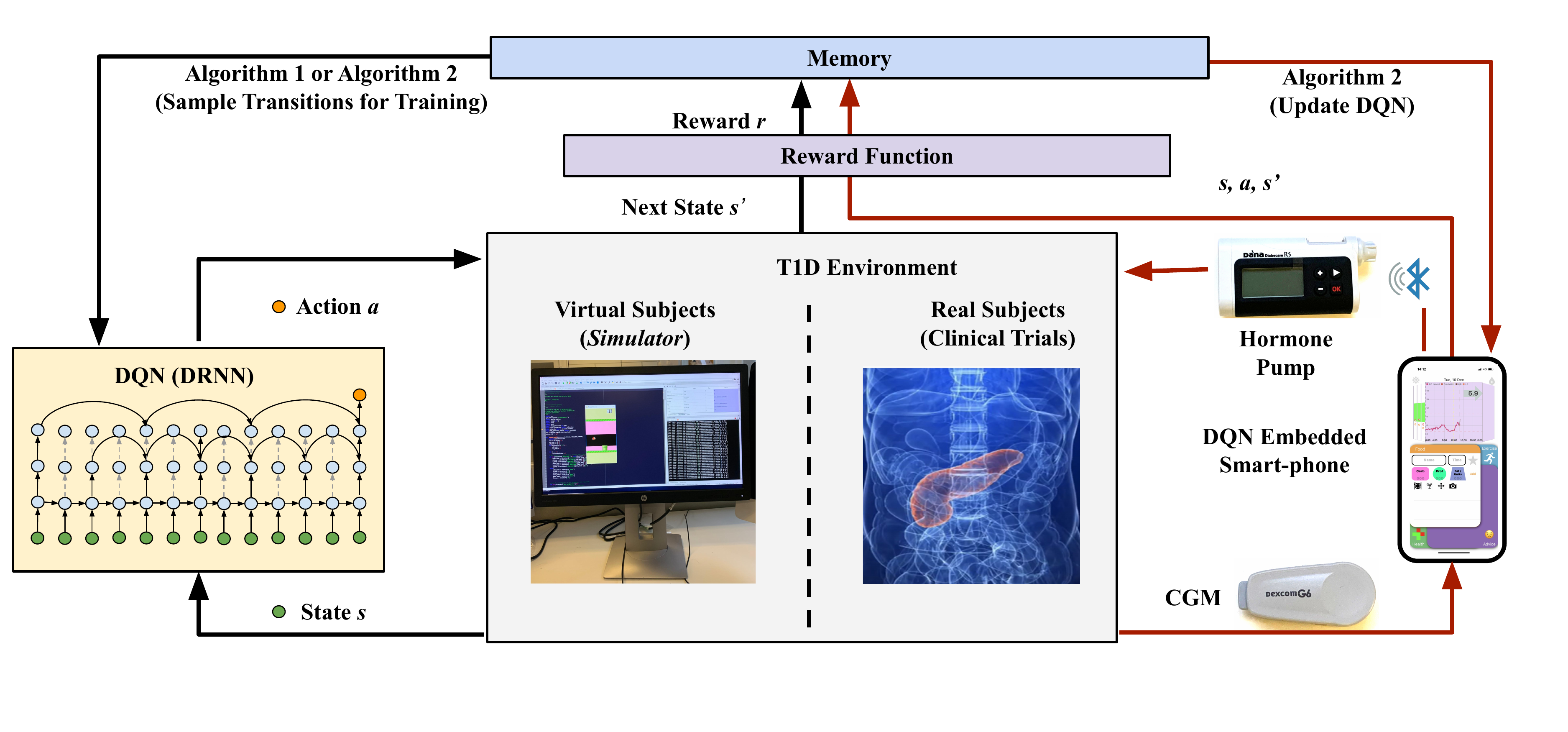} 
\caption{The system architecture to implement deep RL on T1D in the simulator (black arrows) and clinical trials (red arrows).}
\label{fig:system_architecture}
\end{figure*}

Fig. \ref{fig:system_architecture} depicts an overview of the system architecture used to develop the DQN controllers evaluated on the T1D\textit{ in silico} environment and to be potentially used in clinical trials. Algorithm 1 and Algorithm 2 correspond to the two-step learning framework in Section \ref{sec:two-step}.

\subsection{Problem Formulation}
The problem of basal glucose closed-loop control in T1D can be formulated as an infinite-state Markov decision process with noise, which is defined by a tuple $\langle S, \mathcal{P},A, R,\gamma \rangle$ consisting of a state $S$ (i.e., physiological state), a state transition function $\mathcal{P}$, an action $A$ (i.e., insulin and glucagon control actions), a reward function $R$ (i.e., glycemic outcomes), and a discount factor $\gamma \in [0,1]$. The agent in the environment takes an action $a\in A$ at each time step (i.e. each CGM measurement), and then its state $s\in S$ turns into the successor state $s'$ according to $\mathcal{P}$. The policy to select action for given states is denoted by $\pi$. Maximizing the accumulation of expected reward $r_t = R(s_t,a_t)$ at each time step $t$ is the target of RL. An action-value (Q-function) $Q^{\pi}(s,a)$ can be defined to computed this reward:

\begin{equation}
    Q^{\pi}(s,a) = \mathbb{E}[\sum^{\infty}_{t' = t}\gamma^{t'-t}r_{t'}|s_t = s, a_t=a,\pi].
\end{equation}
The optimal action-value function $Q^*(s,a)=\max_{\pi}Q^{\pi}(s,a)$ offers the maximal values, which can be determined by solving the Bellman equation defined by
\begin{equation}
Q^{*}(s,a)  = \mathbb{E}_{s'}\left[ R(s,a) + \gamma \max_{a'}Q^*(s',a')\right] ,
\label{eq:bellman}
\end{equation}
The optimal action-value at the current state $s$ is obtained by selecting the action that maximizes expected return with the optimal $Q^{*}(s',a')$ at the next state $s'$. Although this recursive equation can be estimated by an iterative update, linear and non-linear approximators are commonly used in RL for better generalization \cite{mnih-human-level2015}. In this paper, DQNs are employed to approximate the action-values $Q(s,a;\theta)\approx Q^{*}(s,a)$ where $\theta$ represents the parameters of the neural networks.

\subsubsection{Agent states}
In the closed-loop glucose control problem, we collect the multi-modal data from the control system, as shown in Fig. \ref{fig:system_architecture}, to form a multi-dimensional input vector $D$ to approximate physiological state $S$. Specifically, $D$ comprises the real-time continuous blood glucose levels ${G}$ (mg/dL) measured with a CGM sensor, the carbohydrate estimation of meal ingestion ${M}$ (g) recorder through a smartphone application, and hormone doses delivered by the infusion pumps, including the meal bolus insulin ${B}$, basal insulin $Bas$, and glucagon dose $C$. Thus, we have $D =\{ G,M,I,C\} = [d_{t+1-L}, \cdots,d_t]^\mathbb{T} \in \mathbb{R}^{\mathbb{L}\times 4}$, where $L$ is the length of the time steps vector, $I=B+Bas$ (Unit) represents the sum of meal bolus insulin and basal insulin. The approximated observation $o_t = s_t+e_t$ takes into account the errors or miss-estimations $e_t$ in glucose measurements $G$, carbohydrate meal estimation $M$, and the meal insulin bolus $B$. Here $B$ is computed from $M$ with a standard bolus calculator \cite{schmidt2014bolus}. 
From a deep RL perspective, the problem can be seen as an agent interacting with an environment over sequential time steps. Every five minutes, an observation $o_t$ can be obtained from the environment, and an action $a_t$ can be taken according to the agent's policy.

\subsubsection{Actions}
\label{sec:action}
Following the same framework, we provide two types of delivery strategies for different pump settings. For people with T1D wearing insulin pumps, the action space is defined by modifying the basal insulin rate (BR) as follows: \{suspension of BR, 0.5*BR, BR, 1,5*BR, 2*BR\}. For those wearing dual-hormone pumps, the action space is defined by the following options: \{suspension of BR, 0.5*BR, BR, 1,5*BR, 2*BR, delivering glucagon\}. Note that the value of BR is subject-specific and is known in advance. Based on previous works, we fix glucagon doses to 0.3 \textmu g/kg for all individuals and constraint the total amount of delivered glucagon to a maximum of one mg per day \cite{herrero2017coordinated}.

\subsubsection{Rewards}
\begin{figure}[t]
  \centering
  \includegraphics[width=8cm]{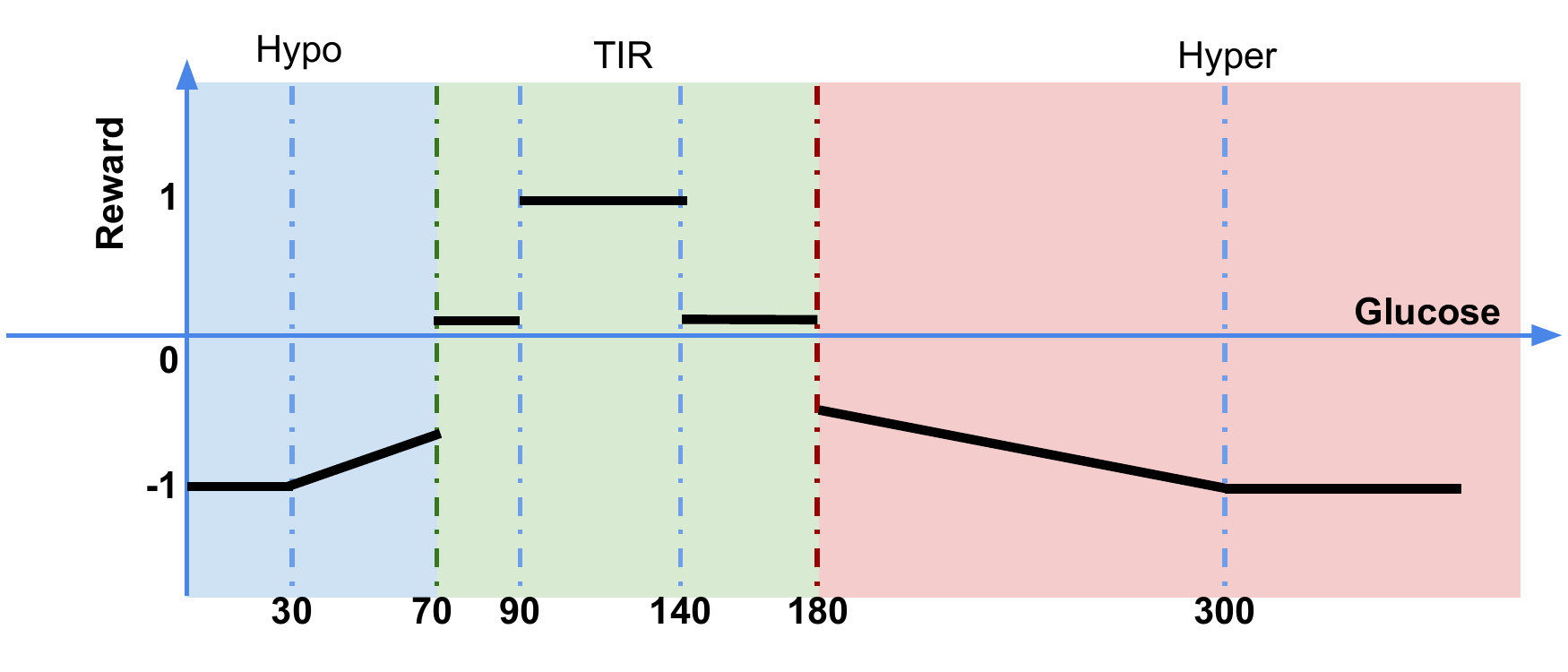}
  \centering
  \caption{Visualization of the employed reward function in terms of the glucose level (mg/dL) in the next state. %
  }
  
  \label{fig:reward}
 \end{figure}

The desired performance of closed-loop glucose control is to maintain BG in a target range of $70$-$180$ mg/dL. By using an empirical approached aiming at maximising time in range (TIR) and minimising hypoglycemia, the following piece-wise reward function was selected.
\begin{equation}
\small
r_t =     \left\{\begin{array}{l}1, \ \ \qquad \qquad \qquad\qquad\qquad\qquad  90\leq G_{t+1}\leq 140\\ 0.1, \ \ \ \qquad 70\leq G_{t+1}< 90 \ \& \  140< G_{t+1}\leq 180 \\ -0.4-(G_{t+1}-180)/200, \ \ \ \ \ 180< G_{t+1}\leq 300 \\ -0.6+(G_{t+1}-70)/100, \ \ \ \ \ \ \ \ \ 30\leq G_{t+1}< 70 \\ -1,  \ \qquad\qquad\qquad\qquad\qquad\qquad \text{else.} \end{array}\right.
\label{eq:reward}
\end{equation}
As depicted in Fig. \ref{fig:reward}, the agent receives a positive reward if the BG level for the next state is in the target range and a negative reward otherwise. If the BG is below 30 mg/dL or above 300 mg/dL, we terminate exploration and restart the simulator. Different evaluated reward functions are presented in Section \ref{sec:reward_functions} of the Appendix.

 \subsection{Two-step Learning Framework}
 \label{sec:two-step}

First, we perform long-term generalized training to obtain a population model for the hormone delivery strategies. We use dilated recurrent neural networks \cite{Chen-DilatedRec2018} for modeling the multi-dimensional time series including glucose levels, hormone doses, and meal intake. Note that other inputs affecting glucose levels, such as physical exercise, could also be considered. To train the model, each basal hormone delivery (at five-minute intervals) is regarded as an action taken by the agent, while the glucose level on the next time step is set to the reward by the criteria of time in range (Equation \ref{eq:reward}). Secondly, by initializing the weights obtained from the population model, we have a model with good initial performance. With a transfer learning process, we individualize the DQNs according to personal characteristics and safety constraints with a small subject-specific data-set. 

\begin{figure}[t]
  \centering
  \includegraphics[width=0.5\textwidth]{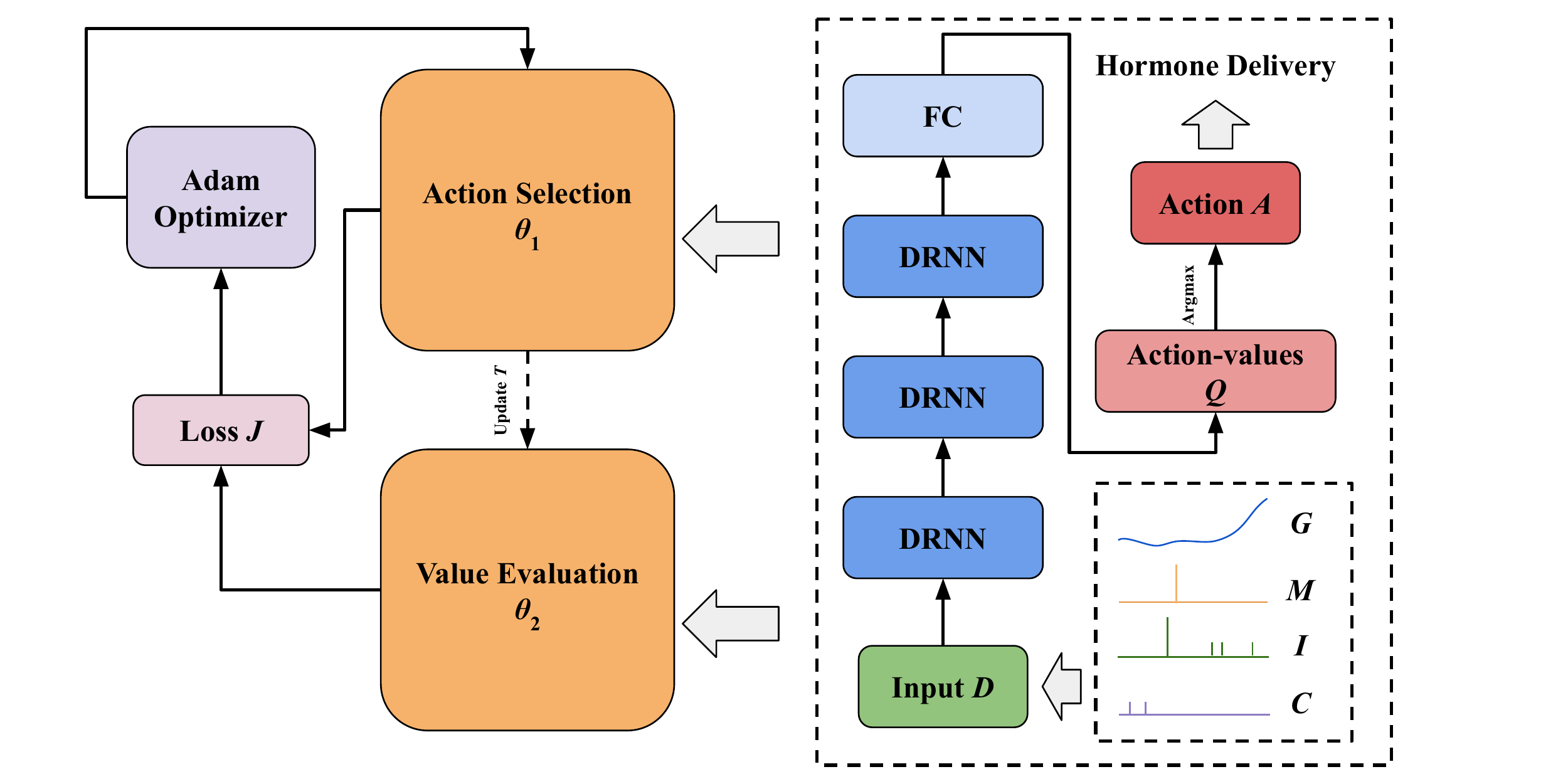}
  \centering
  \caption{The diagram of the propose double DQN. The structure of the neural network is the same for both action selection and value evaluation, which consists of an input layer, a stack of DRNN layers, a fully-connected (FC) layer and output. The input data includes BG series from CGM $G$, meal $M$, insulin $I$ and glucagon $C$. }\label{fig:DQN}
 \end{figure}

During clinical trials, the data for training is usually very limited, thus we aim at fast learning performance. Therefore, we use a double DQN with modified importance sampling to further optimize approximated action values. A state-of-art technique is employed to accelerate learning processes, where prioritized experience replay samples important transitions more frequently \cite{schaul-prioritized2015, hester-deepQ2018}. To avoid overestimating the action values, a double DQN decouples action selection and value evaluation by two separate neural networks \cite{van-deepRL2016}, as shown in Fig. \ref{fig:DQN}. The second step is suitable for a clinical trial setting, where the model is able to adjust itself in a relatively short period of time.

\subsection{Generalized DQN Training}

In the first step, we use the simulator to generate an environment by using the average T1D subject for each one of the virtual cohorts (i.e. adult and adolescent). 
Compared to standard RNNs, DRNNs are preferred as DQNs for learning the delivery strategies. The large receptive field brought by dilation is powerful to extract features from glucose time series, where the dilated skip connection can be represented as 
\begin{equation}
    c_t^{(l)} = f\left( n_t^{(l)}, c_{{t-d}^{(l)}}^{(l)}\right),
\end{equation}
where $c_t^{(l)}$ is the cell in layer $l$ at time $t$, $n_t^{(l)}$ is the input to layer $l$ at time $t$, $d^{(l)}$ denotes the dilation of layer $l$, and $f\left( \cdot \right)$ represents the output function of RNN cells. As shown in Fig. \ref{fig:DQN}, we use three DRNN layers with exponentially increasing dilation, to process the multi-dimensional time-aligned sequence and extract high-level features.. 
Then training is carried out in the simulator with double DQN weights $\theta_1,\theta_2$, where action selections $\theta_1$ and value evaluations $\theta_2$ are obtained from two separate neural networks. According to Equation (\ref{eq:bellman}), the action-selection networks are trained with the loss as
\begin{equation}
\begin{split}
J_{DQ}(Q) &= \mathbb{E}_{(o,a,r,o')\sim\rho}[\left( r + \right. \\ & \ \ \ \  \left. \gamma Q(o',a' ;\theta_2) -Q(o,a;\theta_1) \right)^2],
\end{split}
\end{equation}
where $\rho$ is a mini-batch with transitions $(o,a,r,o')$ sampled from the memory pool, and $a'={\operatorname {arg~max} }_a' Q(o',a';\theta_1) $ is chosen by the action selection DQN in Fig. \ref{fig:DQN}. Thus, the Q-function can be updated as
\begin{equation}
\begin{split}
Q_{\theta_1}(o,a) &\leftarrow Q_{\theta_1}(o,a) + \alpha (r + \\ &\gamma Q_{\theta_2}(o', {\operatorname {arg~max} }_a Q_{\theta_1}(o',a)) -Q_{\theta_1}(o,a)),
\end{split}
\end{equation}
where $\alpha$ is the learning rate, and the weights of $\theta_1$ are copied to $\theta_2$ with a fixed period. We optimize the learning rate by Adam method at each iteration \cite{KingmaB14}. The corresponding pseudo-code is presented in Algorithm 1. 

\begin{algorithm}[ht]
\caption{Generalized DQN Training}
\begin{algorithmic}[1]

\State \textbf {Input:} the environment $E$ with average T1D subject parameters $Is$ provided by the simulator, update period $T_G$, $\varepsilon$-greedy

\State Initialize DQNs with random weights $\theta_1$, $\theta_2$, replay memory $\mathcal{B}$

\For{steps t$\in {1,2,..k}$} 
\State Sample action from $a\sim \pi(Q_{\theta_1},\varepsilon)$, observe $o'$ in $E_{Is}$, calculate $r$, store $(o,a,r,o')$ into $\mathcal{B}$
\EndFor

\Repeat
\State Sample action from $a\sim \pi(Q_{\theta_1},\varepsilon)$, observe $o'$ in $E_{Is}$, calculate $r$, store $(o,a,r,o')$ into $\mathcal{B}$
\State Sample a mini-batch uniformly from $\mathcal{B}$ and calculate loss $J_{DQ}(Q)$ 
\State Perform a gradient descent to update $\theta_1$ 
\State \textbf{if} $t$ \text{mod} $T_G = 0$ \textbf{then} $\theta_2\leftarrow \theta_1$ \textbf{end if} 
\Until{converge}
\end{algorithmic}
\end{algorithm}

For each meal, a standard dose of bolus insulin is delivered, and the agent explores random hormone delivery actions (single or dual) under policy $\pi$ that is $\varepsilon$-greedy with respect to $Q_{\theta_1}$. In this case, random actions are tested with great flexibility and no safety concerns since the simulator is employed for this purpose. 
Note that some human intervention or demonstration at the beginning of the RL process could potentially reduce the training time slightly, but in our case, it was not deemed necessary. At the end of this step, a generalized model consisting of a double DQN with weights $\theta_1$ and $\theta_2$ is obtained.     

\subsection{Personalized DQN Training}

After developing a generalized model, we fine-tune the model by transfer learning with regards to the personal characteristic. We fetch the weights and features from the generalized model, then train the personalized DQNs within a data-set corresponding to a short period of time with safety constraints. We can choose to fine-tune all layers of the generalized model or to retain the weights of some of the earlier layers and only fine-tune a higher-level portion of the network to avoid over-fitting. In experiments, we found that earlier layers contain more generic features (e.g. insulin suspension during the trend of hypoglycemia) that should be useful for all the subjects with T1D. 

Here a method modified from \cite{hester-deepQ2018} is used for calculating the loss of policy-generated data.
Specifically, $J_n(Q)$ has an $n$-step returns ($n$= 12) to propagate values of actions to earlier states $r_t+\gamma r_{i+1}+ \cdots + \gamma^{n-1}r_{i+n-1} + \max_{a}Q(o_{i+n},a)$, and $J_{L_2}(Q)$ is an L2 regularization loss applied to $\theta$ to mitigate over-fitting. Prioritized experience replay samples the transitions with a probability $Pr_i$ proportional to its importance priority \cite{schaul-prioritized2015}, which is computed from previous data and normalized afterwards,
\begin{equation}
    %Pr_i \sim \frac{J_n(Q)_i p_i}{\sum_t{J_n(Q)_t}p_i}
    Pr_i = \frac{p_i^{\alpha}}{\sum_i{p_i^{\alpha}}}, \ \ p_i = |\delta_i|+\epsilon', %\frac{}{\sum_t{J'_n^{\alpha}(Q)_i}}
\end{equation} 
where $\alpha \in [0,1]$ determines the level of using prioritization, $ p_i$ is the priority of transition $i$ calculated from last temporal-difference (TD) error $\delta_i$ and $\epsilon'$ is a small positive constant.
It allows the DQN to more frequently replay transitions with higher TD error. In addition, to ensure that hormones are delivered safely in the clinical trial, constraints $\mathcal{C}$ are applied to the suggested action before execution. Here we use a simple strategy for the safety constraints: suspending basal insulin or glucagon when the current BG level is below 80 mg/dL or over 160 mg/dL, respectively. In practice, the trend and prediction of BG levels can also be used in the safety constraints for early interventions. With proper training of the generalized model and adequate safety constraints, this algorithm can be adopted in a clinical trial setting. The corresponding pseudo-code detailing the algorithm is presented in Algorithm 2.

\begin{algorithm}
\caption{Personalized DQN Training}
\begin{algorithmic}[1]
\State \textbf{Input:} replay memory $\mathcal{B}$ and DQNs weights $\theta'_1$, $\theta'_2$ from generalized training; individual environment $E$, safety constraints $\mathcal{C}$, update period $T_P$, parameter $\lambda_1, \lambda_2$,
\State  Initialize personalized DQNs weights $\theta_1\leftarrow \theta'_1$, $\theta_2\leftarrow \theta'_2$
\State Initialize replay memory $\mathcal{D}$, merging  $\mathcal{B}$ with priorities
\For{steps i$\in {1,2,..N}$} 
\State Sample action from policy $a\sim \pi(Q_{\theta_1})$, 
\State \textbf{if} {$a$ \text{subject to} $\mathcal{C}$ } \textbf{then} execute $a$  \textbf{end if}
\State Observe $(o',r)$ in $E$
\State Store $(o,a,r,o')$ in $\mathcal{D}$, overwriting the samples previously merged from $\mathcal{B}$
\State Sample a mini-batch from $\mathcal{D}$ by modified importance sampling $Pr$ and update the transition priority
\State Calculate loss $J(Q) = J_{DQ}(Q)+\lambda_1 J_n(Q)+\lambda_2 J_{L_2}(Q)$
\State Perform a gradient descent to update $\theta_1$ 
\State \textbf{if} $t$ \text{mod} $T_P = 0$ \textbf{then} $\theta_2\leftarrow \theta_1$ \textbf{end if}  
\EndFor
\end{algorithmic}
\end{algorithm}

\section{Experiments}
\label{sec:Experiments}
Following the architecture evaluation setup depicted in Fig \ref{fig:system_architecture}, we conducted experiments to evaluate, \textit{in silico}, the effectiveness of proposed deep RL framework with the UVA/Padova T1D Simulator \cite{Man-UVA/PADOVA2014}. As stated in Section~\ref{sec:action}, we use two settings of control actions in the proposed deep RL (DRL) algorithm: single-hormone (DRL-SH) and dual-hormone delivery (DRL-DH). Following a transfer learning strategy, we started with a long-term exploration with 1,500 simulated days to obtain a stable generalized model using Algorithm 1, then, we performed personalized training for each individual in the cohort (i.e. adult and adolescent) with 30 simulated days using Algorithm 2. Due to the significant amount of data required, the generalized model is meant to be trained in the simulator, whereas the personalized model training has the potential to be done in a clinical setting. Finally, the personalized models were tested in a period of 90 days.
\linespread{1.2}

\subsection{Experimental Setup}

\subsubsection{\textit{In Silico} environment}

The UVA/Padova T1D simulator provides an interactive environment for the agent to explore and learn the policy. We introduced additional intra-subject variability in the meal protocol scenario and the parameters of the T1D model \cite{herrero2015method}. In particular, we selected four meals as the daily pattern (average cases: 7 am (70 g), 10 am (30 g), 2 pm (110 g), 9 pm (90 g))
with meal-time variability ($STD = 60$ min) and meal-size variability ($CV=10\%$). The meal-duration was set to 15 minutes. A misestimation of carbohydrate amount between $-30\%$ and $+10\%$ with uniform distribution was applied. Variability for meal absorption and carbohydrate bioavailability were set to 30\% and 10\%, respectively. The variability of insulin sensitivity was considered to be 30\% for adult cohort and 20\% for adolescent cohort, which are created by the scenario function in the subjects' own profile. These values of variability were selected based on available physiological knowledge and to achieve the glycemic outcomes commonly observed in such populations when treated with standard therapy \cite{forlenza2018predictive}. 
We saved intra-day and intra-person variability for each subject and used the same scenarios for all the evaluated methods, i.e. same daily events and variability time series, in order to have a fair comparison. 
We utilized the 10 virtual adults and 10 virtual adolescents, plus the corresponding average subjects, for generalized training. 

\subsubsection{Baseline method}
As a baseline method, a low-glucose insulin suspension (LGS) strategy, commonly found in sensor-augmented insulin pumps, was employed \cite{liu2020modular}. LGS systems have been proven to reduce hypoglycemia by suspending basal insulin delivery \cite{Battelino2017Prevention}. For meal bolus calculation, a standard bolus calculator was used \cite{schmidt2014bolus}.

\linespread{1}

\subsection{Results}

To evaluate the performance of the proposed algorithms and compare them against the baseline method, we selected five standard glycemic metrics commonly employed by the diabetes technology community \cite{maahs2016outcome}. These include: percentage time in the glucose target range of $[70,180]$ mg/dL (TIR), percentage time below $ 70$ mg/dL (i.e. hypoglycemia) (Hypo), percentage time above $180$ mg/dL (i.e. hyperglycemia) (Hyper), mean BG levels (Mean), and risk index (RI). Results are expressed by mean values and standard deviations (mean $\pm SD$).

\begin{table*}[!t]

\caption{The testing performance of glucose control on the adult virtual cohort}
\label{tab:adult}
\small
\begin{center}
\begin{threeparttable}    

\begin{tabularx}{0.95\textwidth}{XXXXXX}
\hline
Method   & TIR (\%) & Hypo (\%) & Hyper (\%)  & Mean (mg/dL)  & RI\\ \hline\noalign{\smallskip}
LGS         &$77.55\!\pm \!6.78$ & $2.87\!\pm \!1.38$ & $19.58\!\pm \!5.79$   & $140.78\!\pm \!8.23$   &$2.52\!\pm \!0.89$\\ 
DRL-SH     & $80.94\!\pm\! 7.00^\ast$& $2.06\!\pm\!1.33^\ast$ & $17.00\!\pm\!5.82$     &$140.36\!\pm\!5.98$ &$2.28\!\pm \!0.72$\\ 
DRL-DH    & ${85.55}\!\pm \!{7.33}^{\ast\ast,\dag}$ & ${1.92}\!\pm \!{1.90}^\ast$ &  ${13.81}\!\pm \!{6.67}^{\ast\ast,\dag}$    &${140.12}\!\pm \!{8.13}$ &$2.16\!\pm \!0.65^{\dag}$\\
 \noalign{\smallskip}\hline\noalign{\smallskip}
\end{tabularx}

\begin{tablenotes}
\footnotesize
\item {Symbol $^\ast$ indicates statistical significance ($p\leq 0.05$ ) with with respect to the low-glucose suspension (LGS) and $^\dag$ indicates statistical significance ($p\leq 0.05$) with with respect to the single-hormone DRL (DRL-SH). A double symbol (e.g. $^{\ddagger}$) indicates statistical significance ($p\leq 0.01$).}
\end{tablenotes}
% {\color{red}  
\end{threeparttable}
\end{center}
\end{table*}

\begin{table*}[!t]
\caption{The testing performance of glucose control on the adolescent virtual cohort}
\label{tab:adole}
\small
\begin{center}
\begin{threeparttable}   
\begin{tabularx}{0.95\textwidth}{XXXXXX}

\hline
 Method   & TIR (\%) & Hypo (\%) & Hyper (\%)  & Mean (mg/dL)  & RI\\ \hline\noalign{\smallskip}
 LGS         &$55.50\!\pm \!14.68$ & $6.93\!\pm \!4.69$  & $37.57\!\pm \!11.64$  &$162.15\!\pm \!20.46$  &$4.76\!\pm\!2.70$  \\ 
DRL-SH     & $65.85\!\pm\!16.30^{\ast\ast}$& $5.51\!\pm\!3.37$ & $28.63\!\pm\!14.36^{\ast\ast}$     & $151.18\!\pm\!18.26^{\ast\ast}$ &$3.99\!\pm\!2.43^{\ast\ast}$ \\ 
DRL-DH    & ${78.83}\!\pm \!{6.60}^{\ast\ast,\dag}$ &  ${2.64}\!\pm \!{1.96}^{\ast\ast,\ddagger} $ & ${18.53}\!\pm \!{6.48}^{\ast\ast,\dag}$   &$149.96\!\pm \!8.83^{\ast\ast}$  & $2.94\!\pm\!0.99^{\ast\ast,\ddagger}$ \\
 \noalign{\smallskip}\hline\noalign{\smallskip}
       
\end{tabularx}

\begin{tablenotes}
\footnotesize
\item {Symbol $^\ast$ indicates statistical significance ($p\leq 0.05$ ) with with respect to the low-glucose suspension (LGS) and $^\dag$ indicates statistical significance ($p\leq 0.05$) with with respect to the single-hormone DRL (DRL-SH). A double symbol (e.g. $^{\ddagger}$) indicates statistical significance ($p\leq 0.01$).}
\end{tablenotes}

\end{threeparttable}
\end{center}
\end{table*}

\begin{figure*}[!t]
    \centering

    \begin{subfigure}[b]{0.32\textwidth}
        \centering
         \includegraphics[height=2.2in]{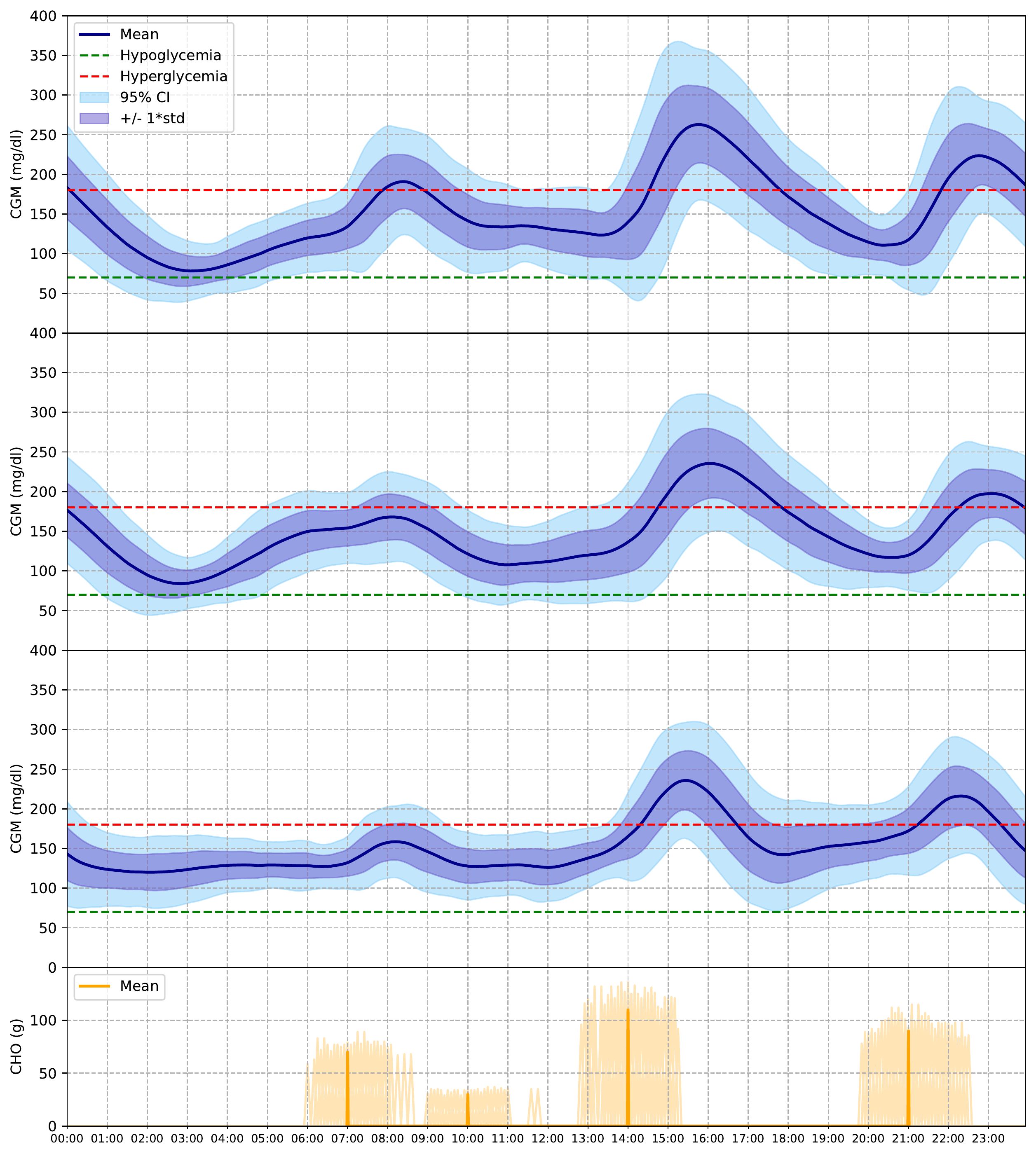}

         \caption{An adult subject simulation}
    \end{subfigure}%
    ~ 
    \begin{subfigure}[b]{0.32\textwidth}
        \centering
         \includegraphics[height=2.2in]{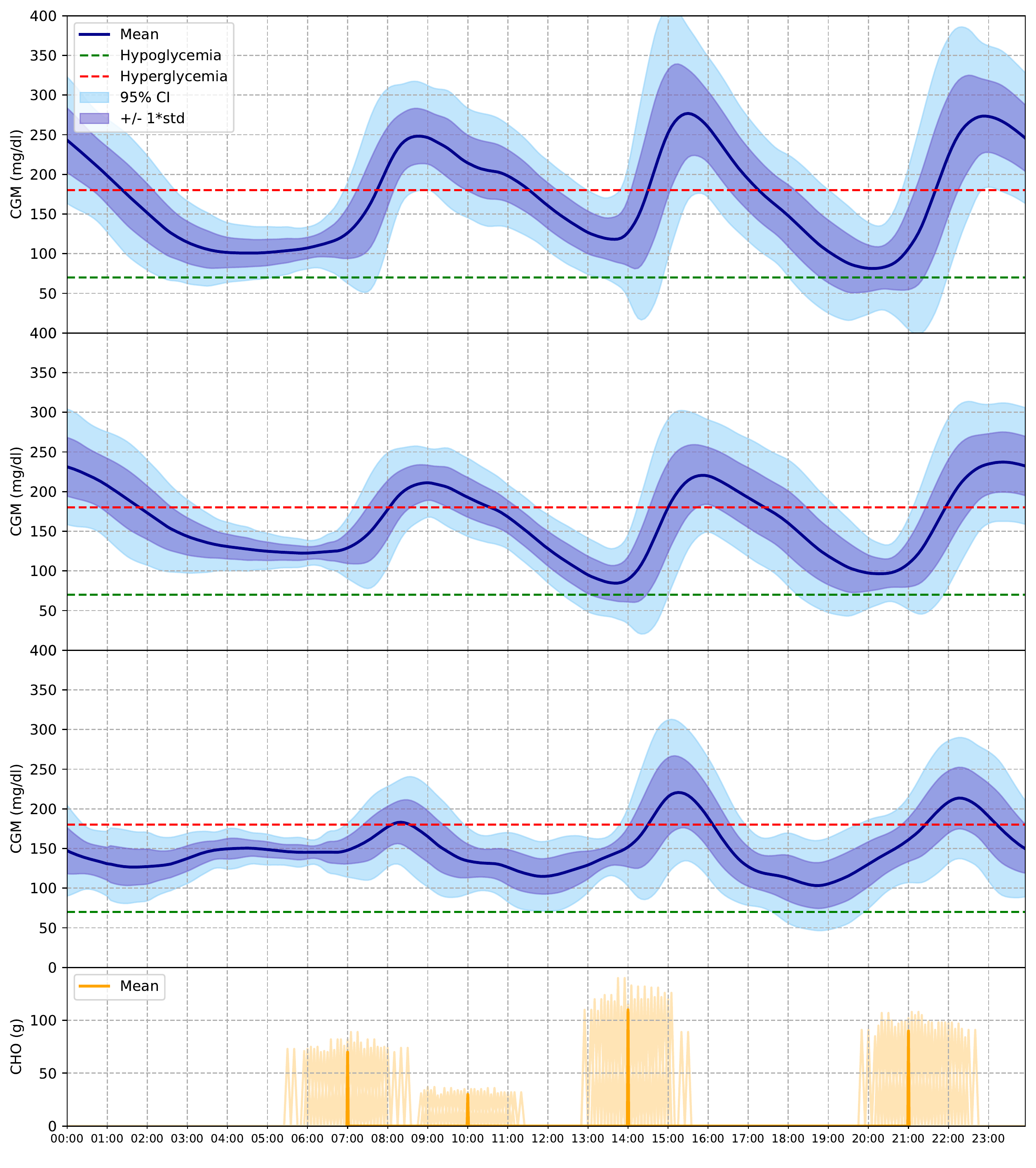}

        \caption{An adolescent subject simulation}
    \end{subfigure}%
        \begin{minipage}[b]{0.33\textwidth}
        \centering
       \begin{subfigure}[b]{\linewidth}
       \centering

	    \includegraphics[height=1.2in]{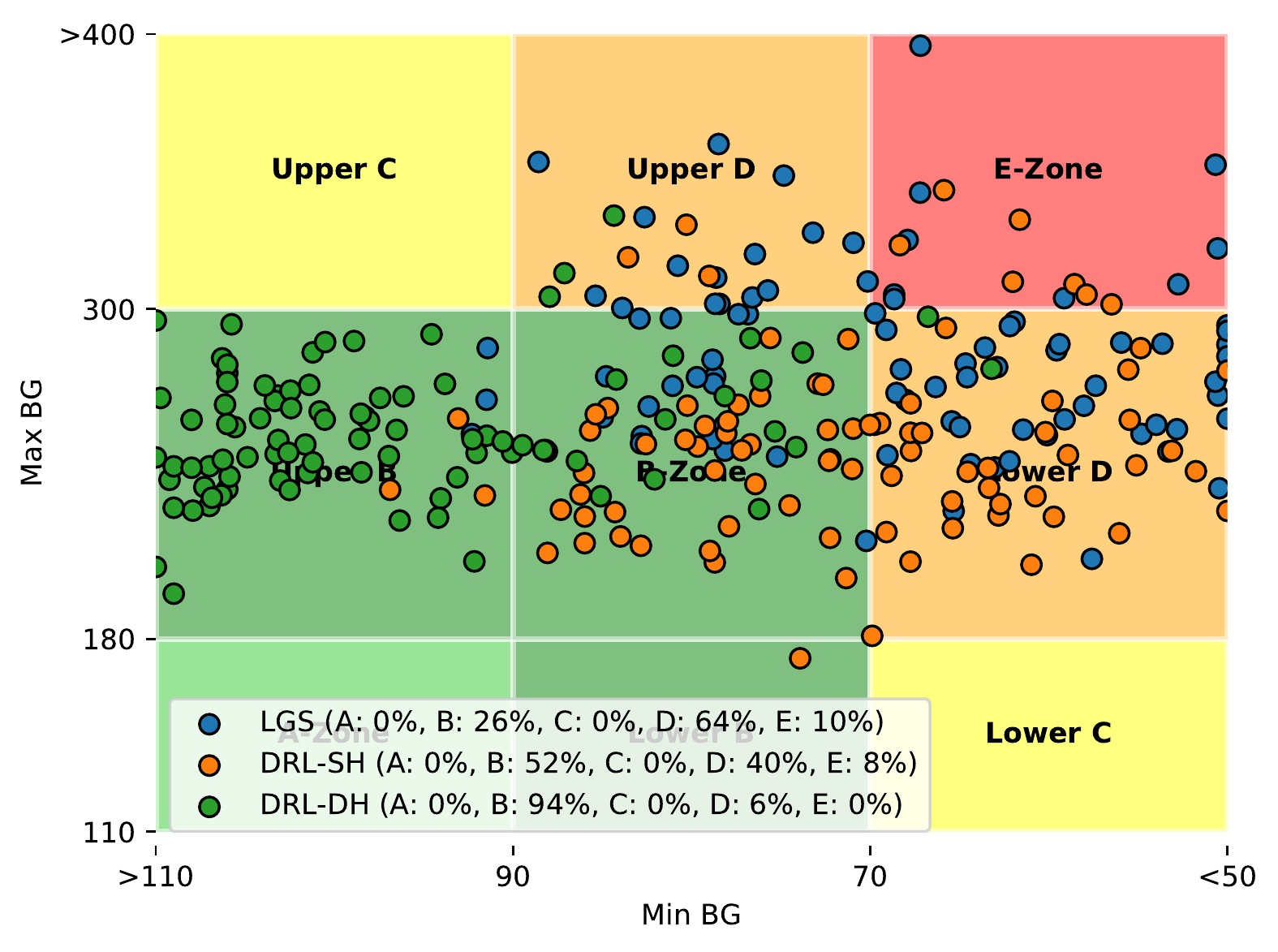}

       \end{subfigure}\\[\baselineskip]
       \begin{subfigure}[b]{\linewidth}
       \centering

	    \includegraphics[height=1.2in]{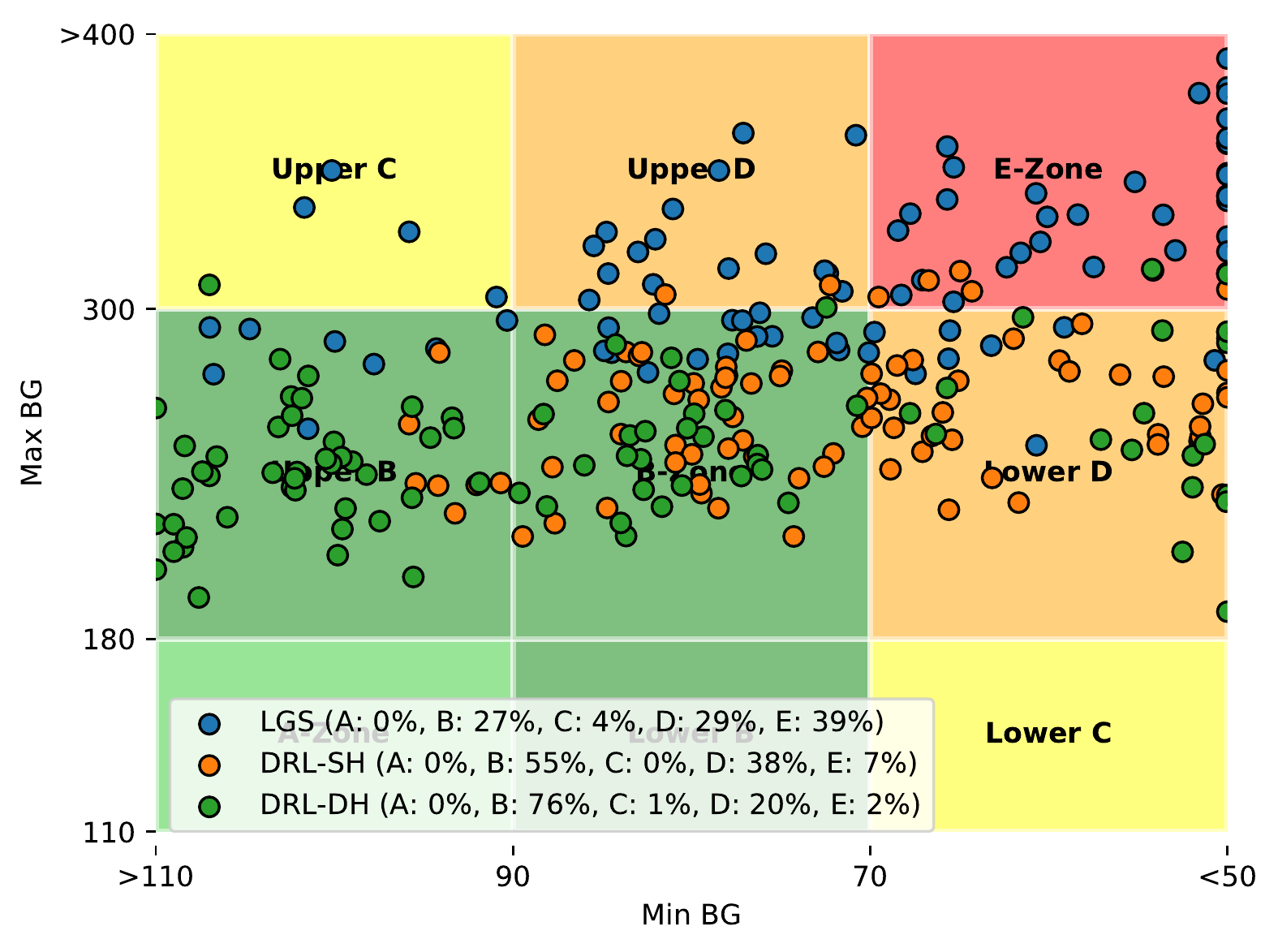}
         \caption{CVGA plots}
       \end{subfigure}
     \end{minipage}

\caption{Visualization of the experiment results for T1D subjects. {\bf (a)} and {\bf (b)}: Performance of the three methods on an adult subject and an adolescent subject over the three-month testing period: (Top-to-bottom) LGS, DRL-SH, DRL-DH, carbohydrate distribution. The average BG levels are shown in solid blue lines, and the hypo/hyperglycemia thresholds are shown in dotted green/red lines. Blue shaded regions show the 95\% confidence interval (CI), and the purple shaded regions indicate the standard deviation. {\bf (c)}: The control variability grid analysis (CVGA) plot for the adult (Top) and the adolescent (Bottom). The blue, orange and green dots represent the LGS, DRL-SH and DRL-DH results, respectively.}
\label{fig:BG_curve_CVGA}
\end{figure*}

Table \ref{tab:adult} and Table \ref{tab:adole} shows the results of the three tested methods evaluated on the adult and adolescent cohorts, respectively. Compared to LGS therapy, both single-hormone and dual-hormone DRL models improve the glucose control performance by reducing hypoglycemia, hyperglycemia and increasing TIR in the two cohorts. Of note, the dual-hormone DRL model significantly increases the mean TIR with a notable decrease of risk index, achieving the best performance. Mean BG levels are maintained in the adult cohort, while the improvement is significant in the adolescent cohort.

For demonstration purposes, Fig. \ref{fig:BG_curve_CVGA} graphically displays the performance of the three evaluated methods for a chosen adult and a chosen adolescent over a three-month testing period. In particular, the glucose profile over 24 hours (mean $\pm SD$) (i.e. ambulatory glucose profile) and the control variability grid analysis (CVGA), a commonly used tool for evaluating closed-loop insulin delivery techniques, were employed \cite{magni2008evaluating}.
Note that the displayed results in Fig. \ref{fig:BG_curve_CVGA} are consistent with the numerical results corresponding to the overall population presented in Tables \ref{tab:adult} and \ref{tab:adole}.
Regarding the CVGA, it is worth noting the significant improvement achieved by DRL-DH when compared to LGS. In particular, the percentage of points in the A+B zones increases from 26\% to 94\% for the adult cohort and from 27\% to 76\% for the adolescent cohort.

\section{Discussion}
\label{sec:Discussion}
\subsection{Comparison with State-of-the-art}

In the presented \textit{in silico} experiments, when compared against a low-glucose insulin suspension technique, the proposed methodology based on deep RL achieves superior performance in terms of glycemic outcomes. Comparing the proposed technique with existing closed-loop insulin delivery techniques, although interesting, is a challenging task due to the difficulty in replicating the testing scenarios and the tuning of the controllers. Hence, a head-to-head comparison has not been performed.
However, although not directly comparable, an informal comparison with existing works in the literature on RL for insulin and glucagon delivery has been done. In \cite{sun-adual2018}, the authors propose an RL-based controller and achieve the adult TIR of $89\%$ on the UVA/Padova simulator, which is close to the performance achieved by our DRL-DH model. Note that in this previous work both basal and bolus insulin delivery are optimized, while in our work only basal insulin delivery is optimized using different variability in the simulator. In \cite{Ngo-ControlOfBl2018}, Ngo and colleges use RL to optimize control parameters in glycemic models without providing comparable TIR results. In a later paper \cite{fox-reinforcement2019}, the authors propose a DQNs algorithm to control insulin delivery and they evaluate it on the previous version of the UVA/Padova simulator. However, no comparable glycemic outcomes are provided. Therefore, our work not only proposes a novel deep RL algorithm for insulin and glucagon delivery but also serves as a benchmark for the future evaluation of other control algorithms. In W3PHIAI-20 workshop~\cite{zhu2020personalized}, we briefly reported some preliminary results corresponding to the dual-hormone delivery configuration. In this paper, we extend this preliminary work by developing a new model for single-hormone delivery, improve the previous algorithms, use more realistic scenarios, and introduce a new baseline method for comparison purposes. To our knowledge, this is the first study that systematically evaluates, \textit{in silico}, a deep RL algorithm to control blood glucose levels with single-hormone and dual-hormone delivery, using the latest T1D simulator (2014 version)~\cite{Man-UVA/PADOVA2014} and additional intra-subject variability.

\subsection{Limitations and Future Work}
Although the DQN models achieved superior control performance \textit{in silico}, clinical validation is still required. There are many uncertainties and perturbations in real-world scenarios, and the main limitation of the simulator is over-estimating the efficacy of glycemic interventions. Therefore, in future work, we consider incorporating more data features that related to glucose dynamics, including physical activities and health conditions. Moreover, there is rapid development in deep RL, and we plan to explore the latest advances in this area, such as model-based RL~\cite{hafner2019dream}, which have the potential to further improve glycemic outcomes and accelerate the training process. Following the proposed setup and framework, it is convenient to implement other deep RL techniques in basal glucose control.

\subsection{Towards Clinical Trials}

In the past years, the technological advances in the field of diabetes technology and mobile phones have increased the connectivity between mobile apps, CGM and insulin pumps. As a result, many researchers have integrated control algorithm (single-hormone and dual-hormone) into apps to automatically administer or recommend hormone delivery and have evaluated them in clinical trials \cite{deshpande2019design,herrero2019bio,lewis2016real,el2017home,castle2018randomized} . 
 
We have developed the deep RL models using TensorFlow, hence it is easy to implement such models on smartphones, or embedded devices, by means of TensorFlow Lite converter. This has been previously done by our group for implementing a DNN model on an app for T1D management \cite{Li-CRNN2019,li2019glunet}. This algorithm has the potential to be continuously be trained and refined by the new incoming data from devices (e.g. CGM, pump, activity monitor) and user input (e.g. meals). Therefore, the algorithm proposed in this work can be implemented in a mobile app without much extra work (Fig.~\ref{fig:system_architecture}).

\section{Conclusion}
\label{sec:Conclusion}
With the aim of overcoming the challenge of blood glucose control in T1D, we propose a novel deep RL algorithm for optimizing basal insulin and glucagon delivery. Dilated RNNs are applied to the structure of double DQNs to develop personalized models through a two-step framework that involves transfer learning. When compared to the baseline method with low-glucose insulin suspension, the proposed methodology significantly improves glycemic outcomes in a virtual adult and adolescent population. This works shows that the proposed approach has the potential to be adopted in a clinical setting.

\section{Acknowledgement}
We would like to thank Chengyuan Liu and Mariam Sarfrazc for their help and assistance. The work is supported by EPSRC EP/P00993X/1 and the President's PhD Scholarship at Imperial College London (UK).

\section{Appendix}

\subsection{Neural Network Selection}
\label{sec:DQN_selection}

\begin{figure}[!t]
    \centering

    \begin{subfigure}[b]{1\columnwidth}
        \centering
         \includegraphics[width=0.8\columnwidth]{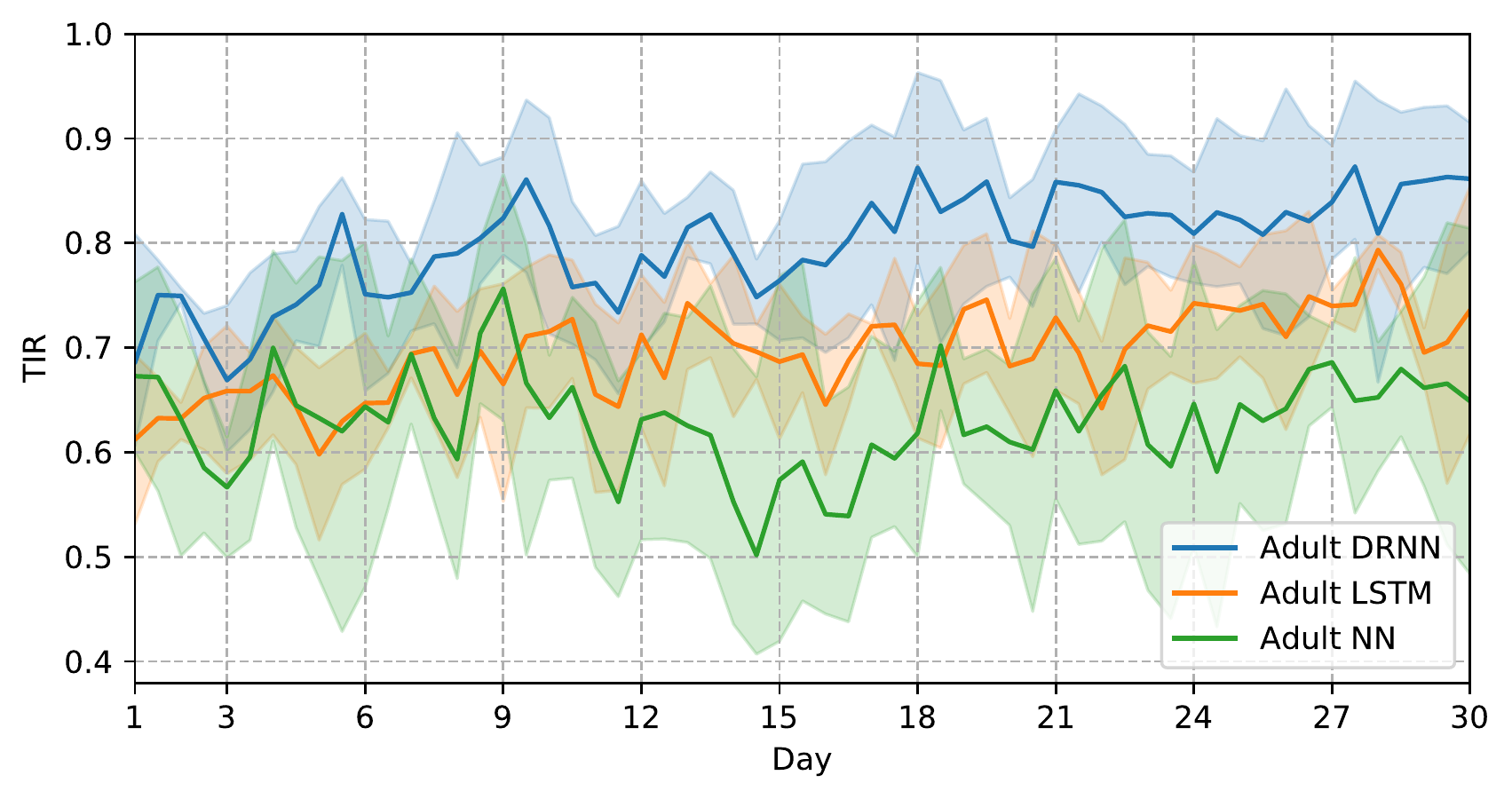}

         \caption{Adult Cohort}
    \end{subfigure}%

    \begin{subfigure}[b]{1\columnwidth}
        \centering
         \includegraphics[width = 0.8\columnwidth]{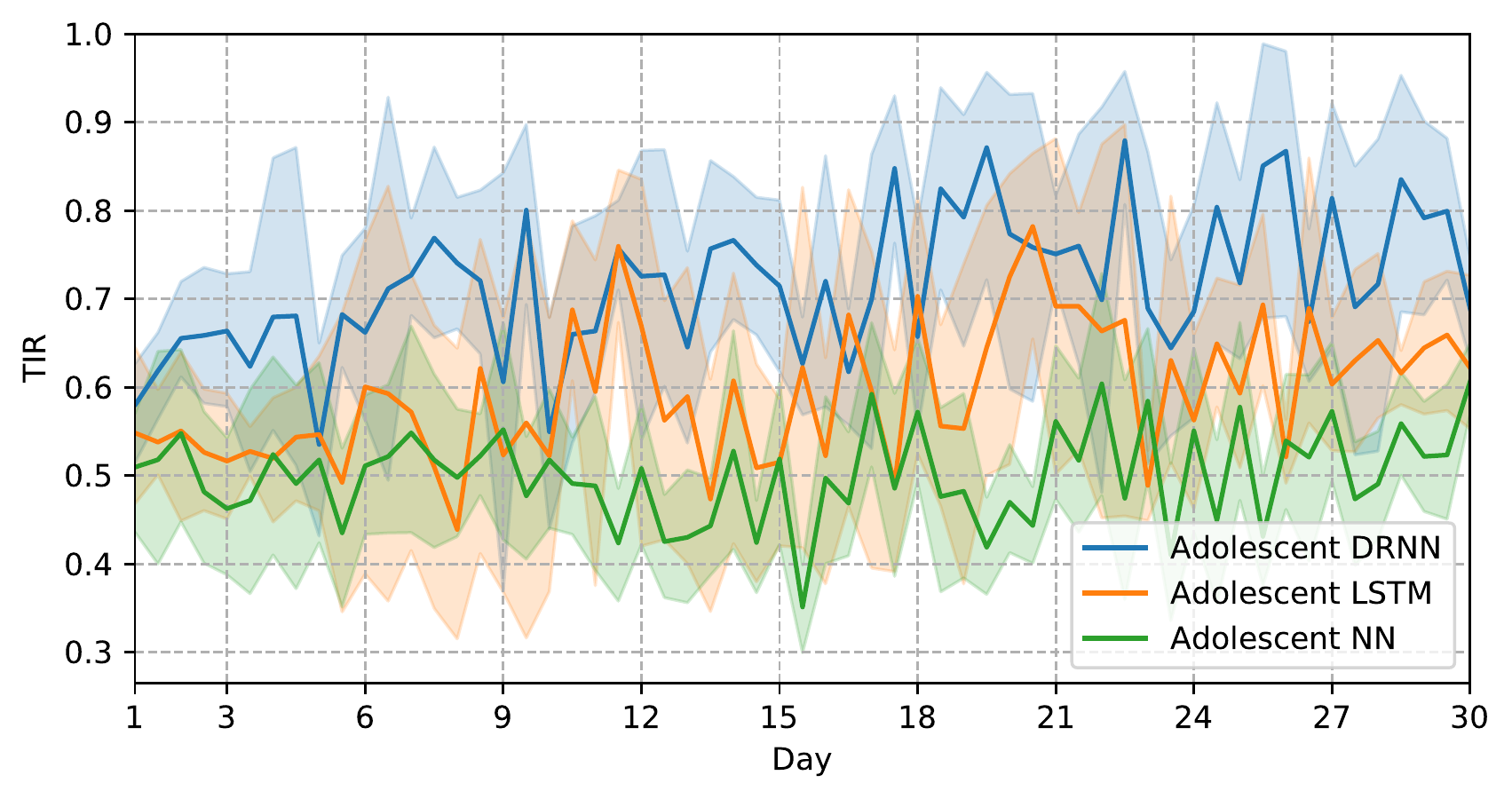}

        \caption{Adolescent cohort}
    \end{subfigure}%

\caption{ TIR results (mean, $95\%$ CI) corresponding to DRL-DH during the personalized training for the adult and adolescent cohorts. The blue, orange and green lines show the results of DRNN, LSTM and NN models, respectively.}
\label{fig:DQN_selection}
\end{figure}

Fig. \ref{fig:DQN_selection} shows the TIR results achieved with the different neural network architectures that we evaluated in the experiments. Considering that the input data is a multi-dimensional time-aligned sequence, we assumed that an RNN-based model would be a good candidate to map the multiple-step historical data. Therefore, we explored conventional long short-term memory (LSTM), NNs with five fully-connected layers and DRNNs as the potential structure of DQNs. The LSTM architecture has recently achieved great success in time-aligned tasks, but in our case, it obtains lower TIR results than the DRNN. NNs are commonly used in DQNs as a basic structure. However, the NN curve in Fig. \ref{fig:DQN_selection} shows large variability and lower mean TIR. Less variability indicates a better capability to account for within-subject variability. Thus, the NN structure was discarded. Regarding the DRNNs, the generalized model achieves a good initial performance at the beginning of personalized training. In addition, the DRNN curve has a positive trend and small variability, which indicates its effectiveness at adjusting the models for a specific subject through a short period of time. Finally, DRNN prediction models ranked top in Blood Glucose Level Prediction Challenge in 2018 \cite{Chen-DilatedRec2018}. Therefore, DRNNs was selected as the DQNs for this work.

\subsection{Reward Function}
\label{sec:reward_functions}

\begin{table}[ht]
\centering
\resizebox{\columnwidth}{!}{\begin{tabular}{|c|c|c|c|c|}
\rowcolor[HTML]{343434} 
{\color[HTML]{FFFFFF} \textit{\begin{tabular}[c]{@{}c@{}}GL Range \\ (mg/dL) \end{tabular}}} & {\color[HTML]{FFFFFF} \begin{tabular}[c]{@{}c@{}}Reward \\ Scheme 1\end{tabular}} & {\color[HTML]{FFFFFF} \begin{tabular}[c]{@{}c@{}}Reward \\ Scheme 2\end{tabular}} & {\color[HTML]{FFFFFF} \begin{tabular}[c]{@{}c@{}}Reward \\ Scheme 3\end{tabular}} & {\color[HTML]{FFFFFF} \begin{tabular}[c]{@{}c@{}}Reward \\ Scheme 4\end{tabular}} \\
\cellcolor[HTML]{EFEFEF}{\color[HTML]{343434} \textit{0-30}}                                        & -10                                                                               & -1                                                                                & -1                                                                                & -1                                                                                \\
\cellcolor[HTML]{EFEFEF}{\color[HTML]{343434} \textit{30-70}}                                       & -1                                                                                & -0.5                                                                              &  $-0.5+\frac{GL-70}{80}$                                                                & $-0.6+\frac{GL-70}{100}$                                                              \\
\cellcolor[HTML]{EFEFEF}{\color[HTML]{343434} \textit{70-90}}                                                & +0.1                                                                              & +0.1                                                                              & +0.1                                                                              & +0.1                                                                              \\
\cellcolor[HTML]{EFEFEF}{\color[HTML]{343434} \textit{90-140}}                                      & +1                                                                                & +1                                                                                & +1                                                                                & +1                                                                                \\
\cellcolor[HTML]{EFEFEF}{\color[HTML]{343434} 140-180}                                              & +0.1                                                                              & +0.1                                                                              & +0.1                                                                              & +0.1                                                                              \\
\cellcolor[HTML]{EFEFEF}{\color[HTML]{343434} \textit{180-300}}                                     & -1                                                                                & -0.5                                                                              & $-0.5-\frac{GL-180}{240}$                                                              & $-0.4-\frac{GL-180}{200}$                                                            \\
\cellcolor[HTML]{EFEFEF}{\color[HTML]{343434} \textit{300+}}                                        & -10                                                                               & -1                                                                            & -1                                                                                & -1                                                                                \\
\rowcolor[HTML]{343434} 
{\color[HTML]{FFFFFF} TIR Score (\%)}                                                                   & {\color[HTML]{FFFFFF} 75}                                                         & {\color[HTML]{FFFFFF} 88}                                                         & {\color[HTML]{FFFFFF} 86}                                                         & {\color[HTML]{FFFFFF} 93}                                                        
\end{tabular}}
~
\caption{Reward functions and corresponding scores}
\label{reward_scores}
\end{table}

Crafting reward functions for RL models is one of the most crucial factors determining the model performance.
Note that this performance evaluation is only applicable when the TIR score converges to a fixed value through the training stage for the adult with Algorithm 1 and dual-hormone. Different reward schemes tested, together with their corresponding model scores, can be seen in Table~\ref{reward_scores}. In experiments, we started with a piece-wise step function referred to as Reward Scheme 1, then we constrained the reward function range within [-1, 1] to improve stability (Reward Scheme 2). Afterwards, we introduced slopes into the reward function to make the agent's response to glucose changes smoother (Reward Scheme 2). Note that the agent faces an increasing penalty as glucose moves up in the hyperglycemic range, or down in the hypoglycemic range. Finally, with minor adjustments on the slopes, the best TIR score was obtained by Reward Scheme 4, where a TIR of 93\% can be achieved after 1.3 million training steps.

\subsection{Hyper-parameters}
In Table \ref{tab:hyperparameters}, we list the hyper-parameters that have been used in this work. For each parameter, we performed limited tuning based on the state-of-art DQN~\cite{hessel2018rainbow}.
All the parameters are identical across all the virtual subjects.
\begin{table}[ht]
\centering
\begin{tabular}{l|r}
Parameter                               & Value                 \\ \hline
Exploration before learning  $k$            & 2000        \\
Generalized network update period $T_G$               & 1000        \\

Generalized DQN $\varepsilon$-greedy    & 0.5$\rightarrow$0.01  \\
Personalized network update period $T_P$               & 100        \\
Discount factor $\gamma$ & 0.9                   \\

Adam learning rate                      & $1\times10^{-5}$      \\
Batch size                              & 32                    \\
Number of time steps $L$                         & 12                    \\
Replay buff size $\mathcal{B}$          & 5000         \\
Prioritization exponent $\alpha$        & 0.3                   \\
Importance-sampling exponent $\beta$    & 0.4$\rightarrow$1.0   \\
Multi-step return $\lambda _1$                & 0.1 \\
L2 regularization $\lambda_2$            & $1\times10^{-5}$ \\
Cell type                           & Vanilla RNN      \\ 
DRNN dilation                            & {[}1, 2, 4{]}    \\
Hidden nodes of DRNN layers         & [32, 64, 128]   
\end{tabular}
\caption{List of hyperparameters}
\label{tab:hyperparameters}
\end{table}

\iffalse
comments
\fi

\end{document}